\documentclass[reprint,superscriptaddress,amsmath,twocolumn,secnumarabic,amssymb, nobibnotes, aps, prd]{revtex4-2}

\usepackage{hyperref}
\usepackage{amsmath}
\usepackage{amssymb}
\usepackage{graphicx}
\usepackage{float}
\usepackage{svg}

\hypersetup{hypertex=true,colorlinks=true,linkcolor=blue,anchorcolor=blue,citecolor=blue,urlcolor=blue}

\begin{document}
\title{Manipulation of annular electron beams in plasmas}
\author{Yangchun Liu}
\affiliation{Institute for Fusion Theory and Simulation, School of Physics, Zhejiang University, Hangzhou, 310058, China}
\author{Dong Wu}
\email{dwu.phys@sjtu.edu.cn}
\affiliation{Key Laboratory for Laser Plasmas and School of Physics and Astronomy,and Collaborative Innovation Center of IFSA (CICIFSA),Shanghai Jiao Tong University, Shanghai, 200240, China}
\author{Tianyi Liang}
\affiliation{Institute for Fusion Theory and Simulation, School of Physics, Zhejiang University, Hangzhou, 310058, China}
\author{Zhengmao Sheng}
\email{zmsheng@zju.edu.cn}
\affiliation{Institute for Fusion Theory and Simulation, School of Physics, Zhejiang University, Hangzhou, 310058, China}
\author{Xiantu He}
\affiliation{Institute for Fusion Theory and Simulation, School of Physics, Zhejiang University, Hangzhou, 310058, China}

\date{\today}
\begin{abstract}
   The annular electron beam has significant practical potential in high-energy physics and condensed matter physics, which can be used to edge-enhancement electron imaging, collimation of antiprotons in conventional linear accelerators, acceleration of positively particles like positrons, structured X-ray generation and manipulation of nanomaterials. The quality of an annular electron beam depends on its energy, flux and topology. In this article, we study the transport characteristics of annular electron beam in a plasma medium and propose a scheme to modify it. According to our theory and full three-dimensional LAPINS simulations, we have found that the self-generated magnetic field focuses the incident annular electron beam, enabling the adjustment of its annular width (AW). Besides, the annular electron beam, endowed with orbital angular momentum (OAM), exhibits contrasting transport characteristics compared to an electron beam devoid of OAM. The former requires an external magnetic field to ensure stable transportation in the plasma. Under the influence of this magnetic field, the radius of the annular electron beam oscillates periodically, with the direction of change whether increasing or decreasing dependent on the field's strength. In this case, the radius of annular electron beam will be affected by the external magnetic field and allows for the simultaneous adjustment of its radius and AW, significantly broadening its application range.
\end{abstract}
\maketitle
\section{Introduction}
   With the development of ultra-short and ultra-intense laser technology, the laser intensity including Laguerre-Gaussian (LG) laser has been greatly improved. Now LG laser with intensity as high as $6.3\times 10^{19}\ \mathrm{W/cm^2}$ can be generated in the laboratory through a high-reflectivity phase mirror \cite{WangWP2020PRL} and can accelerate the electron beam to GeV levels \cite{Turner2019PRL}. LG laser has a hollow structure and carries orbital angular momentum (OAM) \cite{Allen1992PRA}, which makes it possible to manipulate micron-sized particles and nanoparticles \cite{Ashkin1970PRL,Andersen2006PRL,Padgett2011NPh}. Besides, LG laser has important applications in super-resolution microscopy \cite{Jesacher2005PRL,Jack2009PRL}, communication \cite{Wang2012NPh,Nenad2013Science,Chen2018PRL}, quantum computing \cite{Molina2007NP,Gibson2004OE}, gravitational wave detection \cite{Mours2006CQG} and plasma acceleration \cite{WangWP2020PRL,Vieira2014PRL}. The interaction between LG laser and plasma can produce annular electron beam \cite{Wang2021HEDP,Maitrallain2022NJP,Pollock2015PRL,WangWP2020PRL,Wang2021HEDP,Wangscrep2015}, which is expected to have novel applications such as edge-enhancement electron imaging \cite{Zhang2017PRL}, collimation of antiprotons in conventional linear accelerators \cite{Stancari2011PRL}, acceleration of positively particles \cite{Jain2015PRL}, structured X-ray generation \cite{Zhao2016PRL} and manipulation of nanomaterials \cite{Lloyd2017RMP}.
\par
   The annular width (AW), which represents the distance between the outer and inner surfaces, and the radius of an annular electron beam are critical factors that significantly influence its range of applications. For instance, in auger electron spectroscopy and electron beam lithography, the AW of the annular electron beam is closely associated with its resolution \cite{Balamuniappan2019IJMPA}. In the context of accelerating positrons with an annular electron beam, the radius of the annular electron beam plays a crucial role in influencing the intensity of the accelerating electric field \cite{Jain2015PRL}. Moreover, the radius of the annular electron beam will also impact the manipulation of nanoparticles \cite{Verbeeck2013AM}. In the previous work, many schemes of generating annular electron beams have been put forward. The simplest way is to use a high voltage pulse generator \cite{Li2008RSI}, but the intensity and energy of the annular electron beam generated by this method are relatively low. The interaction between a laser and plasma can yield a significant quantity of high-energy electrons. Consequently, the scheme of using LG laser which with annular structure to generate annular electron beam is proposed. J. Vieira et al. find nonlinear wakefields driven by LG laser pulses can lead to hollow electron self-injection and positron acceleration \cite{Vieira2014PRL}. Wang and Jiang et al. successfully employed an LG laser to generate annular electron beams with OAM \cite{Jiang2021HPLSE,Wangscrep2015}. Zhao et al. successfully produced a quasi-monoenergetic annular electron beam with an average energy range of $200\sim400\ \mathrm{MeV}$ in the laser wake field accelerator, and thus produced a high-flux X-Ray \cite{Zhao2016PRL}. However, in previous research on annular electron beams, the emphasis was largely on their generation and applications, with minimal investigation into methods for modifying these beams.
\par
   In this article, we studied the transport characteristics of annular electron beams in plasma mediums by the rigid beam model and full three-dimensional (3D) LAPINS simulations \cite{Wu2023POP}. We found that the azimuthal magnetic field which can focus the incident annular electron beam will be generated during its transport in the plasma medium. Moreover, considering that certain schemes for generating annular electron beams will carry OAM \cite{Wang2021HEDP,Wangscrep2015}, our theory and simulations indicate that the possession of OAM by the annular electron beam has minimal impact on the strength of the azimuthal magnetic field. However, we note distinct transport characteristics of the annular electron beam with OAM compared to those without OAM. Specifically, the former necessitates an external axial magnetic field to ensure its stable transportation in the plasma medium. Under the influence of this magnetic field, the radius of the annular electron beam can oscillate periodically, with the direction of change whether increasing or decreasing dependent on the field's strength. In this case, the radius of the annular electron beam will be affected by the external magnetic field and allows for the simultaneous adjustment of its radius and AW, significantly broadening its application range.

\section{Theoretical Model}
   The transport of an annular electron beam within a solid target can be computed using the rigid beam model, as employed by J. R. Davies \cite{Davies2003PRE}. To simplify, we introduce the current neutralization model. This model is valid under the following two conditions. The first one is that the beam duration time is much longer than the electron plasma period $2\pi/\omega_{pe}$, where $\omega_{pe}$ is the electron plasma frequency. The second is that the beam radius is much larger than the plasma electron skin depth $c/\omega_{pe}$, where $c$ is the light speed \cite{Kaganovich2008PoP,Kaganovich2001PoP,Berdanier2015PoP}. According to the generalized Ohm's law, the approximate electric field in plasma is
   \begin{equation}
   	\boldsymbol{E}\approx \eta \boldsymbol{J}_e-\frac{1}{en_e}\nabla p_e,
   	\label{eq1}
   \end{equation}
   where $\eta$ is the resistivity of background plasma, $\boldsymbol{J}_e$ is the return current density of electrons, $e$ is the elementary charge, $n_e$ and $p_e$ are the density and the thermal pressure of plasma electrons. Additionally, we disregard the movement of ions given their substantially greater mass compared to electrons. According to the current neutralization condition, we have $\boldsymbol{J}_e\approx -\boldsymbol{J}_b$, where $\boldsymbol{J}_b$ is the current density of the electron beam. For simplicity, we use
   \begin{equation}
   	n_b=n_{b0}\exp \left[ -\left( r-r_0 \right) ^2/R^2 \right]
   	\label{eq2}
   \end{equation}
   in cylindrical coordinates to represent the annular electron beam. Assuming that the background plasma density and temperature are uniformly distributed, and the annular electron beam density is much smaller than the background plasma electron density. The electric field can be approximated as
   \begin{equation}
   	\boldsymbol{E}\approx \eta \boldsymbol{J}_e\approx - \eta\boldsymbol{J}_b=\eta e n_b\boldsymbol{v}_b,
   	\label{eq3}
   \end{equation}
   where $\boldsymbol{v}_b$ is velocity of the electron beam. When the influence of the stopping power of the target on the electron beam density is ignored and only Ohmic heating is considered, according to the rigid beam model, the temperature change of the background plasma satisfies the following equation after the time $\Delta t$,
   \begin{equation}
   	C\Delta T=\eta \mathrm{J}_{b}^{2}\Delta t,
   	\label{eq4}
   \end{equation}
   where $C=3n_ek/2$ is the heat capacity of plasma. Eq.\ (\ref{eq4}) can also be written as
   \begin{equation}
   	\frac{\partial T}{\partial t}=\frac{2\eta \mathrm{J}_{b}^{2}}{3kn_e}.
    \label{eq5}
   \end{equation}
   According to Spitzer resistivity formulation, we have
   \begin{equation}
   	\eta=\eta_0(T/T_0)^{-3/2},
   	\label{eq6}
   \end{equation}
   where $\eta_0$ and $T_0$ is initial resistivity and temperature, respectively. For simplicity, we define $\tau=t-z/v_b$, which gives the length of time that the beam has been passing a given point. Combing Eqs.\  (\ref{eq5}) and (\ref{eq6}), we find
   \begin{equation}
   	T=T_0\left( 1+\Omega \right) ^{2/5},
   	\label{eq7}
   \end{equation}
   where $\Omega =5\eta _0\mathrm{J}_{\mathrm{b}}^{2}\tau/3n_ekT_0$ means the ratio of Ohmic heating energy to plasma electrons thermal energy. For $\Omega\ll 1$, the plasma temperature is slightly increased, which corresponds to weak heating, and for $\Omega \gg 1$ the plasma temperature is significantly increased, which corresponds to strong heating \cite{Ning2021JPP}. Combing Eqs.\ (\ref{eq6}) and (\ref{eq2}), we have
   \begin{equation}
   	\boldsymbol{E}=-\eta _0\left( 1+\Omega \right) ^{-3/5}\boldsymbol{J}_b.
   	\label{eq8}
   \end{equation}
   According to Faraday's law, we can deduce the transverse magnetic field caused by the propagation of the annular electron beam in the plasma medium. We have
   \begin{align}
   	 	B_{\theta}=&-\frac{3n_ekT_0}{J_b R^2} (r-r_0) \notag \\
   	 	&\times \left[\frac{1}{5}\left( 1+\Omega \right) ^{2/5}+\frac{4}{5}\left( 1+\Omega \right) ^{-3/5}-1 \right].
   	\label{eq9}
   \end{align}
   When the pulse time of the electron beam is relatively short and the background temperature is not exceedingly low, $\Omega$ is generally less than $1$. It means that the value in the bracket of Eq.\ (\ref{eq9}) is less than 0. For $r < r_0$, $B_{\theta} > 0$, electrons are subjected to outward Lorentz force. For $r > r_0$, $B_{\theta} < 0$, electrons are subjected to inward Lorentz force. This causes the annular electron beam to be subjected to a force that decrease the AW when it is transported in the plasma. We can use this principle to control the AW of the annular electron beam when it is transported in the target.

   	Additionally, the focusing condition can be calculated using the motion equation of the electron beam in the transverse direction. By analyzing the force on the electron beam in the transverse direction, we have
   	\begin{equation}
   		n_b\frac{dp_{br}}{dt}=-\frac{dp_{th}}{dr}+J_b B_{\theta}-en_b E_r.
   		\label{eq10}
   	\end{equation}
	Here, $p_{br}$ represents the transverse momentum of the electron beam, and $p_{th} = n_bkT_b$ denotes the thermal pressure of the electron beam. The transverse electric field $E_r$ can be obtained from Eq.\ (\ref{eq1}). When the electron beam density is much lower than the background plasma density, the effect of the electric field is significantly weaker than that of thermal pressure and magnetic pressure. This is because of $n_b \ll n_e, v_e \ll v_b$. Therefore, the focusing conditions of the electron beam are
	\begin{equation}
		\left| J_bB_{\theta} \right|>\left| -\frac{dp_{th}}{dr} \right|.
		\label{eq11}
	\end{equation}
	In that case of weak heating, $\Omega \ll 1$, we have
		\begin{equation}
		B_{\theta}\approx \frac{2\eta _0J_b\tau}{R^2}\left( r-r_0 \right).
		\label{eq12}
	\end{equation}
	Regardless of changes in the electron beam's temperature, the focusing condition of the annular electron beam becomes
	\begin{equation}
		\frac{v_bJ}{kT_b}>\frac{1}{e\eta _0\tau}.
	\end{equation}
	Therefore, the main factors affecting the focusing of the annular electron beam are its current intensity and temperature. Additionally, due to the relationship between background plasma density and resistivity, the background density also influences the focusing characteristics.

   \par
   \begin{figure}[!t]
   	\centering
   	\includegraphics[scale=0.55]{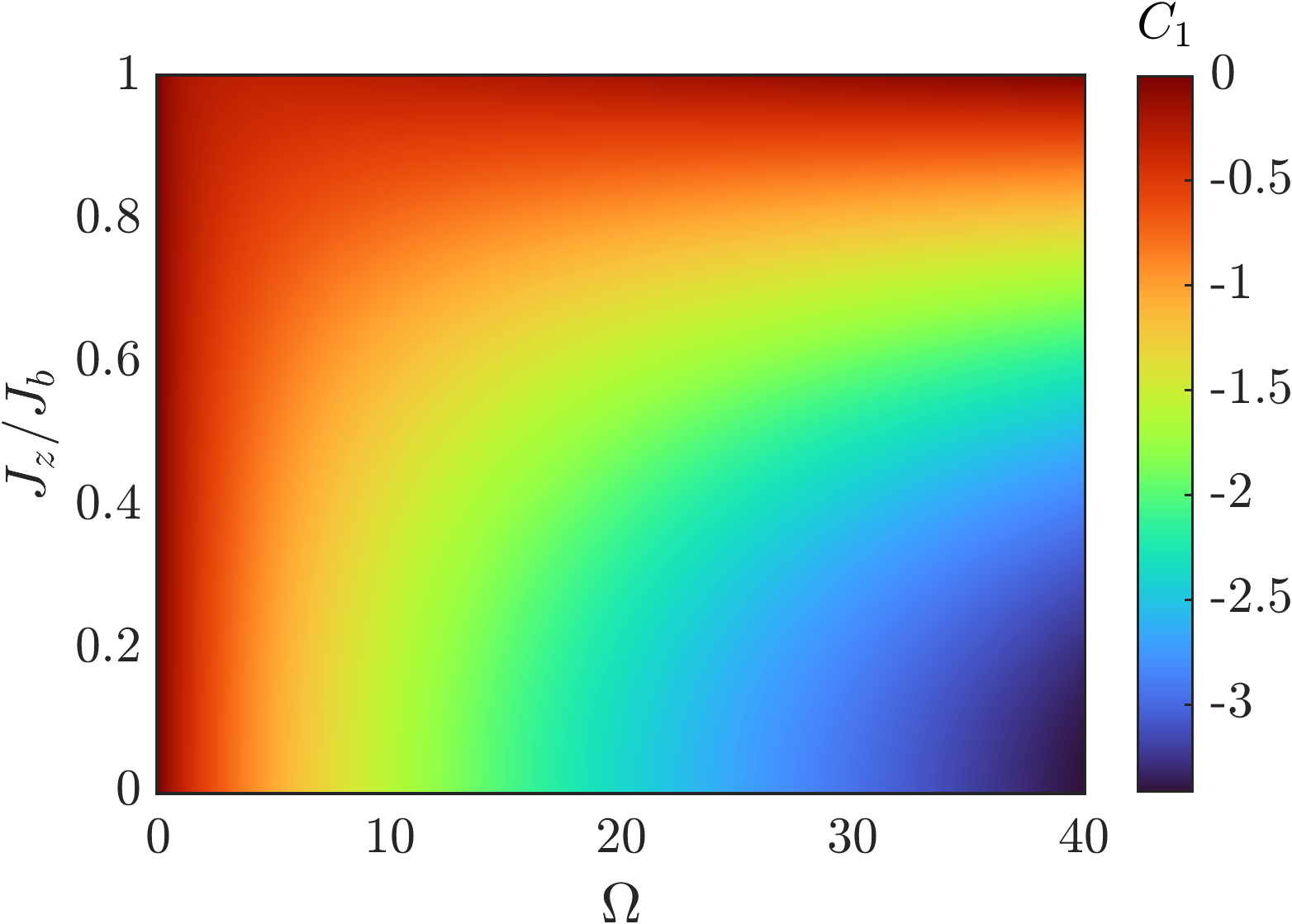} 	
   	\caption{$C_1$ versus $\Omega$ and $J_z/J_b$.}
   	\label{fig1}
   \end{figure}
   Additionally, in certain schemes for generating annular electron beams, the beam exhibits not only forward velocity, but also carries OAM \cite{Wang2021HEDP,Wangscrep2015}. In this scenario, electrons possess not only longitudinal velocity but also velocity in the azimuthal direction. At this point, the velocity of the electron beam can be expressed as $\boldsymbol{v}_b=v_z\boldsymbol{e}_z+v_\theta\boldsymbol{e}_\theta$, where $\boldsymbol{e}_z$ represents the direction of incidence of the electron beam and $\boldsymbol{e}_\theta$ represents the azimuthal direction. Therefore, the azimuthal magnetic field becomes
   \begin{widetext}
    \begin{equation}
    	\begin{split}    		    		\boldsymbol{B}_{\theta}=&\frac{dJ_z}{dr}\frac{3n_ekT_0}{2{J_b}^2}\left [ \frac{6}{5}\left( 1+\Omega \right) ^{2/5}\frac{{J_z}^2}{{J_b}^2}+\frac{4}{5}\left( 1+\Omega \right) ^{-3/5}\frac{{J_z}^2}{{J_b}^2}-2\frac{{J_z}^2}{{J_b}^2}-\left( 1+\Omega \right) ^{2/5}+1 \right ]+ \\&\frac{dJ_{\theta}}{dr}\frac{n_ekT_0}{{J_b}^2}\frac{J_{\theta}J_z}{{\mathrm{J}_b}^2}\left [ \frac{9}{5}\left( 1+\Omega \right) ^{2/5}+\frac{6}{5}\left( 1+\Omega \right) ^{-3/5}-3 \right ],   	
    	\end{split}
    \label{eq10}
    \end{equation}
   \end{widetext}
   where $J_z=-en_bv_z$ represents the current in the longitudinal direction and $J_\theta=-en_bv_\theta$ represents the current in the  azimuthal orientation direction. In order to analyze the physical meaning of the terms in Eq.\ (\ref{eq10}), we define the quantity in the first square brackets of Eq.\ (\ref{eq10}) as $C_1$, and the quantity second square brackets as $C_2$. We can readily find that $C_2$ is always greater than $0$. Moreover, $C_1$ decreases with the increase of $\Omega$ and increases with the increase of $J_z/J_b$, as shown in the Fig.\ \ref{fig1}. Besides, we can easily find $C_1$ is always a negative value in the case of weak heating ($\Omega<1$). We assume that the longitudinal velocity of the electron beam is constant and the electron beam carry OAM has a constant angular velocity $\omega$, then we have $J_\theta=-e n_b\omega r$. According to Eq.\ (\ref{eq10}), it can be found that OAM will decrease the azimuthal magnetic field.
   \par
\begin{figure}[!t]
	\centering
	\includegraphics[scale=0.55]{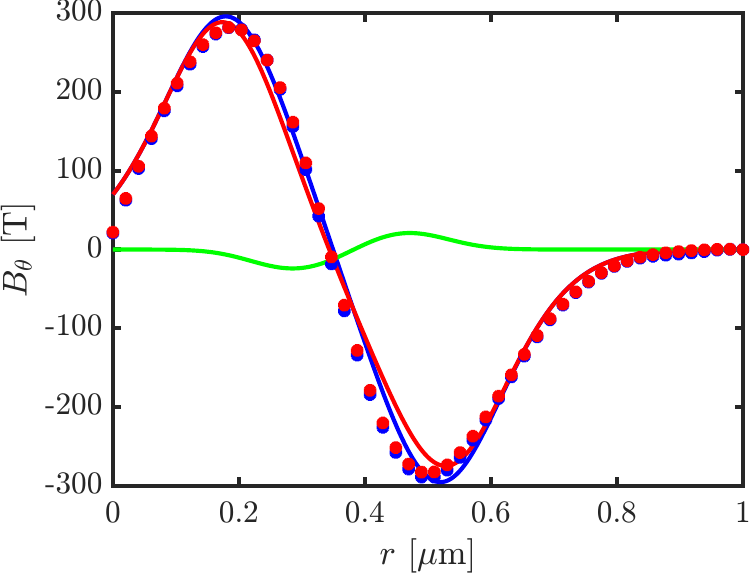} 	
	\caption{Distribution of azimuthal magnetic field in theoretical and simulation results. The red solid line and dotted line show the theoretical and simulation results without OAM, and the blue solid line and dotted line show the theoretical and simulation results with OAM. The green solid line is the value corresponding to the second term in Eq. (\ref{eq10}), which indicates the influence of OAM on the azimuthal magnetic field.}
	\label{fig2}
\end{figure}

   From the above formula, we can see that OAM of the electron beam has a certain inhibitory effect on the azimuthal magnetic field. However, quantitatively analyzing the impact of OAM on the azimuthal magnetic field directly from the above formula is challenging. Hence, we opt to introduce a set of specific parameters and conduct comparative analyses of the results obtained from 3D numerical simulations. For simplicity, the temperature of annular electron beam is set to $1\ \mathrm{eV}$ and its density can be expressed as $n_b=n_{b0}\exp [ -( r-r_0) ^2/R^2 ]$, where $n_{b0}=0.5 n_c$, $r_0=0.35\ \mathrm{\mu m}$, $R=0.2\ \mathrm{\mu m}$ and $n_c=1.11\times 10^{21}\ \mathrm{cm^{-3}}$. For the case without OAM, the initial momentum of the electron beam is $p_z = 6.0m_e c$, where $m_e$ is the electron rest mass. The background plasma density is $n_p=120n_c$ and its initail temperature is $T_e=200\ \mathrm{eV}$. We bring these above data into the formulas and calculate the magnitude of the azimuthal magnetic field for $\tau = 15T_0$, the result is drawn in Fig.\ \ref{fig2} (blue solid line), where $\tau$ is the time that the beam has been passing the diagnostic position and $T_0 = 3.33\ \mathrm{fs}$. According to the above data, we can calculate that the electron plasma period $2\pi/\omega_{pe}$ is about $0.3\ \mathrm{fs}$ and the plasma electron skin depth $c/\omega_{pe}$ is about $0.0146\ \mathrm{\mu m}$, which satisfies the prerequisites of the current neutralization model. For the case that the electron beam carries OAM, we set the initial momentum of the electron beam as $p_b =p_\theta \boldsymbol{e}_\theta+p_z\boldsymbol{e}_z$ and we take $p_{\theta} = 2rm_ec/\lambda$, where $\lambda=1\ \mathrm{\mu m}$ is the unit of normalized length. Other properties of annular electron beam and background plasma remain consistent with the situation without OAM. The theoretical results have been drawn in Fig.\ \ref{fig2} (red solid line). Besides, we also plot the second term in Eq.\ (\ref{eq10}), it represents the influence of OAM on the azimuthal magnetic field (green solid line). Based on this result, we observe that OAM will marginally decrease the azimuthal magnetic field.
   In the following section, we will present our 3D simulation results and compare them with the theoretical findings.

\begin{figure}[!t]
	\centering
	\includegraphics[scale=0.34]{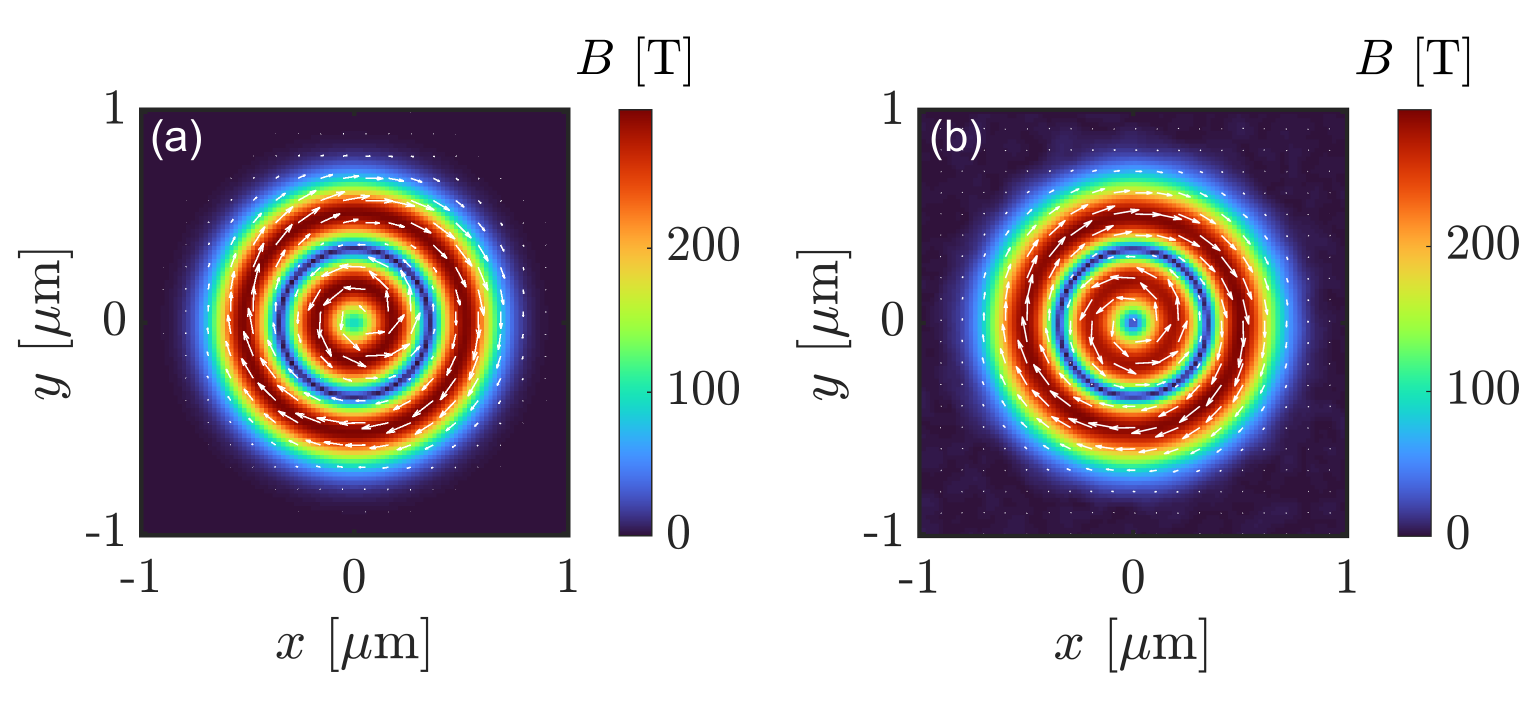} 	
	\caption{(a) and (b) are the theoretical and simulated transverse magnetic field strength for $\tau=15T_0$ in the case without OAM, respectively, where the arrow indicates the direction of the magnetic field. Unless otherwise stated (also in following figures), the time corresponding to the data in the figures is $\tau=15T_0$.}
	\label{fig3}
\end{figure}

\begin{figure}[!t]
	\centering
	\includegraphics[scale=0.32]{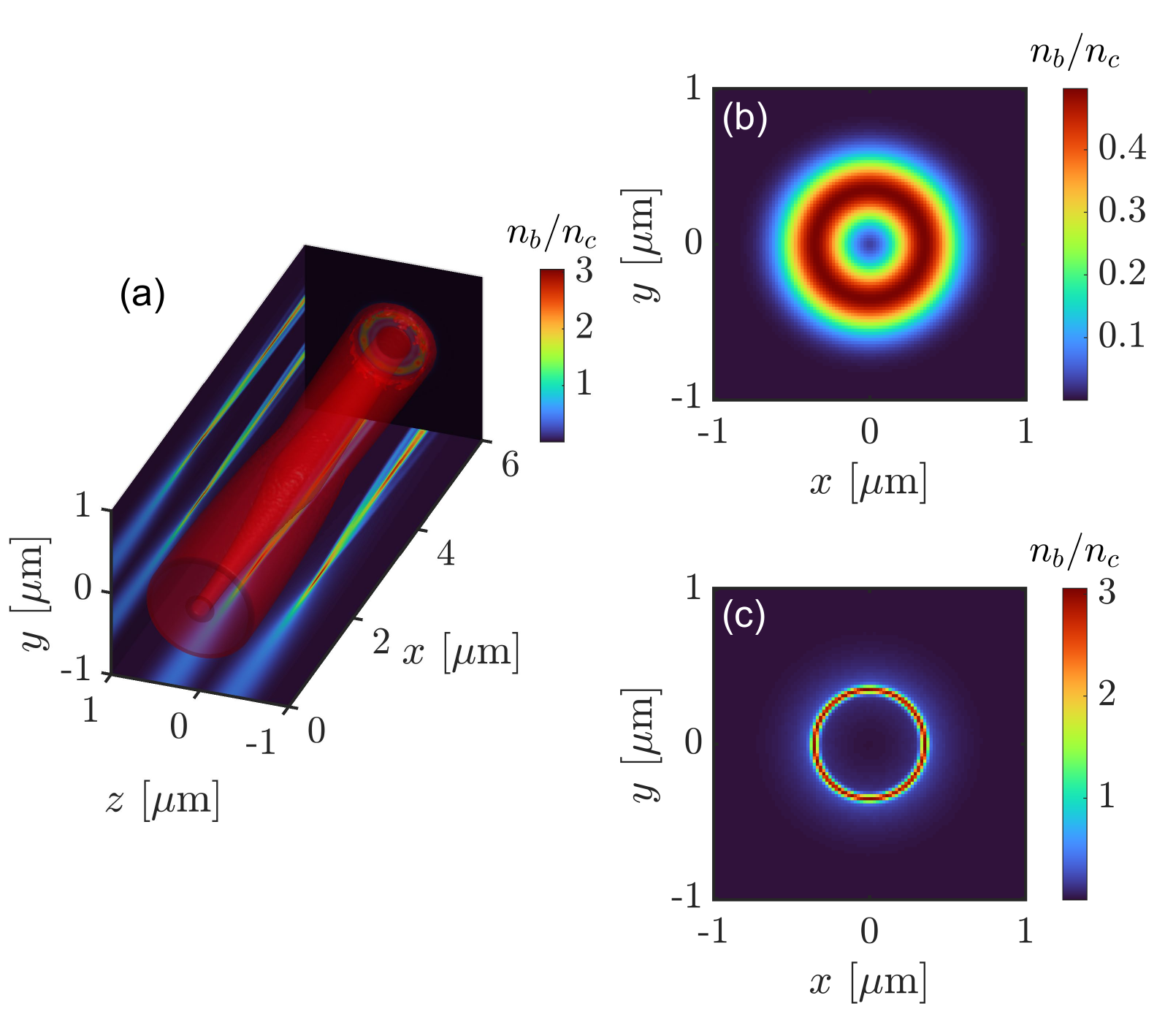} 	
	\caption{Density distribution of annular electron beam in different cross sections for the case without OAM. (a) The red area is an isosurface with $n_b=0.1n_c$. The images on the coordinate axis correspond to the slice data of $x=0\ \mathrm{\mu m}$ and $y=0\ \mathrm{\mu m}$ respectively. (b) and (c) is $n_b$ in the $x,y$ plane at $z=0$ and $2.8\ \mathrm{\mu m}$ respectively.}
	\label{fig4}
\end{figure}

\begin{figure}[!t]
	\centering
	\includegraphics[scale=0.27]{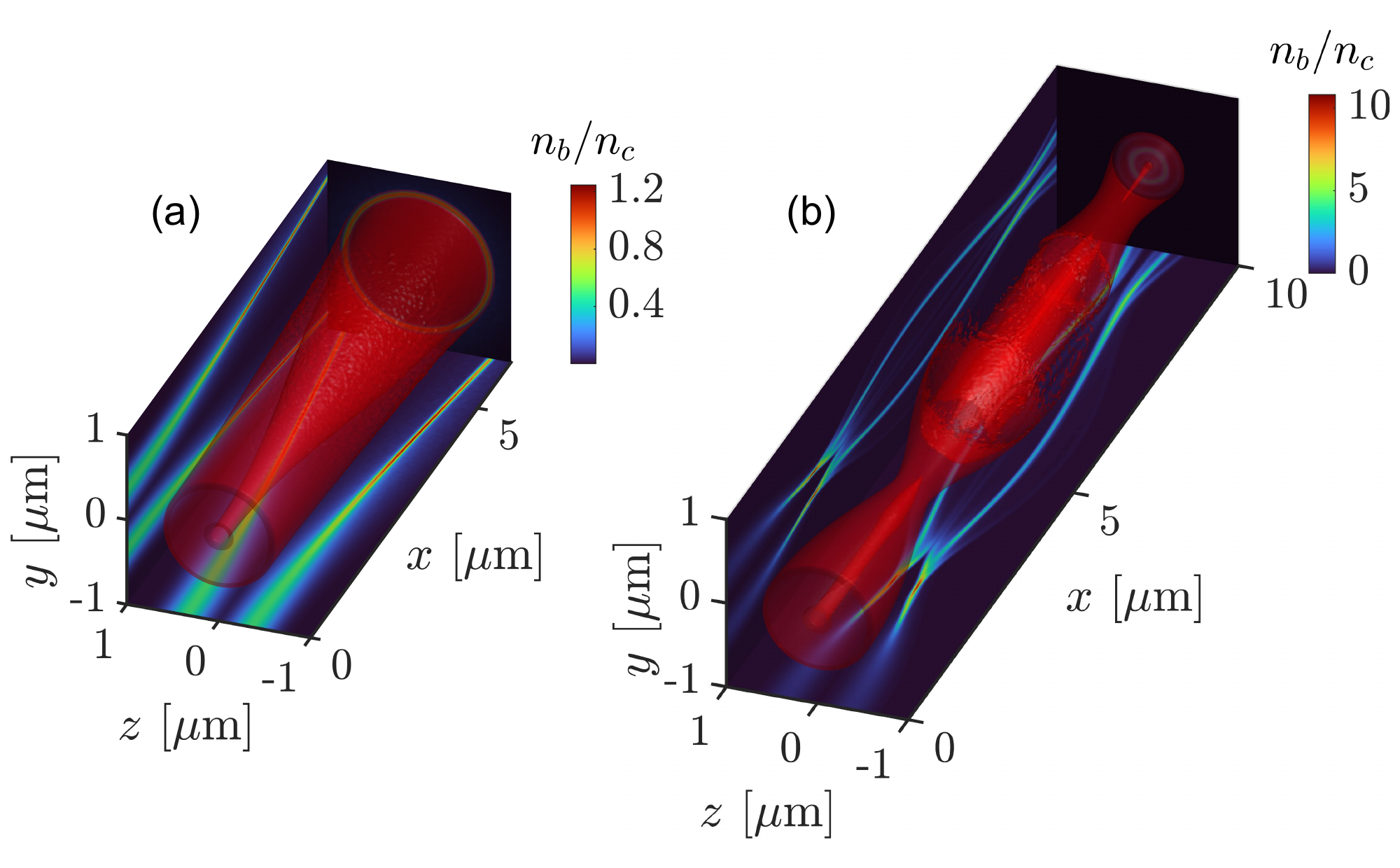} 	
	\caption{(a) and (b) are isosurfaces with $n_b=0.1n_c$ for the OAM case without and with the longitudinal magnetic field, respectively. The images on the coordinate axis correspond to the slice data of $x=0\ \mathrm{\mu m}$ and $y=0\ \mathrm{\mu m}$ respectively.}
	\label{fig5}
\end{figure}

\begin{figure}[!t]
	\centering
	\includegraphics[scale=0.25]{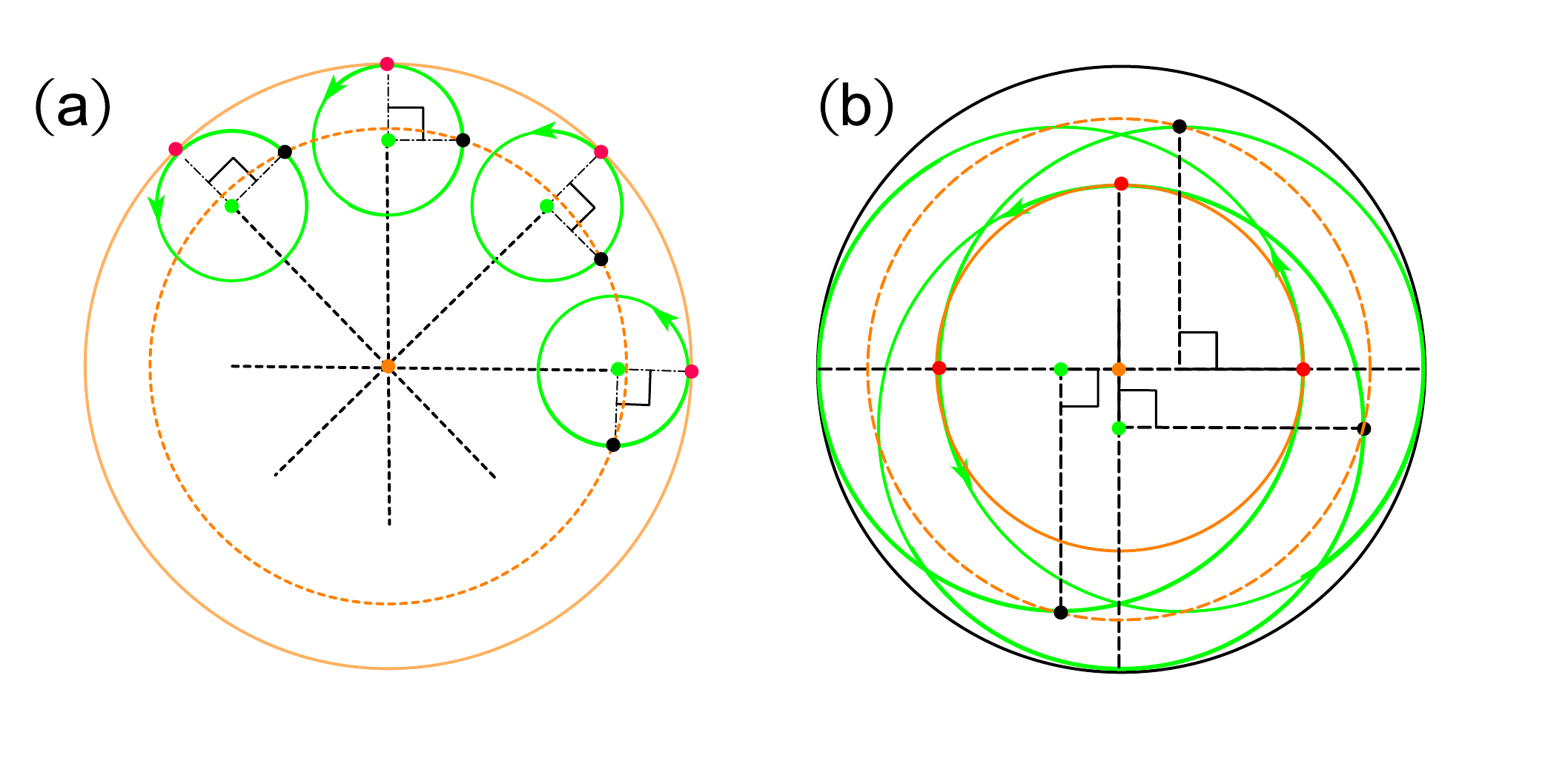} 	
	\caption{(a) and (b) are schematic diagrams of electron motion of in the case $B_z=1\times 10^4\ \mathrm{T}$ and $r_c>r$, respectively. The orange solid and dotted lines represent the outline of the annular electron beam at electron cyclotron phases of $0$ and $3 \pi/2$, respectively. The green solid line traces the path of the electron's cyclotron motion, while the black dotted line extends the electron's cyclotron radius relative to the center of the annular electron beam. The red dots and black dots represent the positions of electrons at cyclotron phases of $0$ and $3\pi/2$, respectively, while the green arrows indicate the direction of the electrons' cyclotron motion. The black solid line represents the maximum radius value during the periodic oscillation of the annular electron beam. The contour of the annular electron beam remains annular. }
	\label{fig6}
\end{figure}

\begin{figure*}
	\centering
	\includegraphics[scale=0.5]{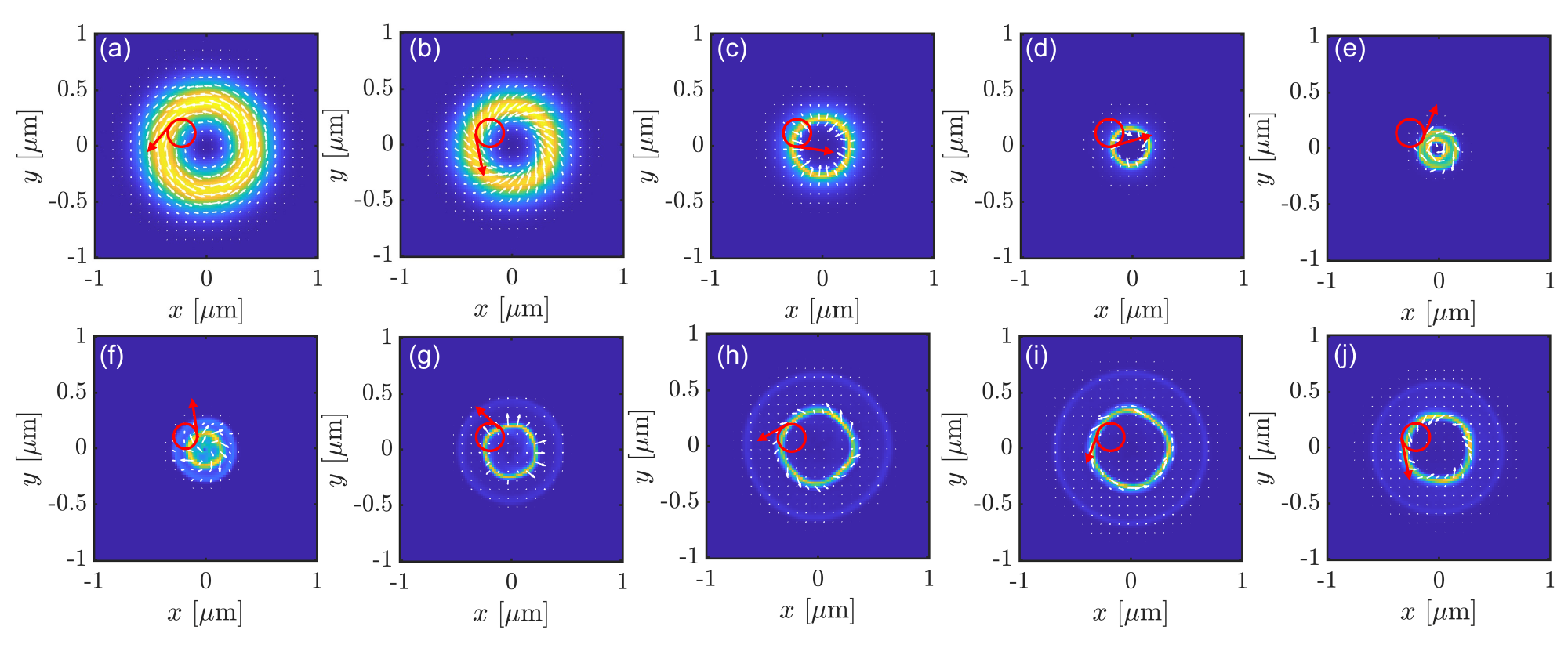} 	
	\caption{Transverse motion of annular electron beam. Arrow direction indicates the transverse movement direction of electrons, in which the white arrows are directly drawn by the current, and the red arrows are the schematic diagram of the transverse movement direction of electrons drawn according to the white arrows. (a)-(j) is the $x,y$ plane at $z=0,\ 1.00,\ 1.70,\ 2.30,\ 3.00,\ 3.50,\ 4.20,\ 5.20,\ 5.90$ and $7.00\ \mathrm{\mu m}$ respectively.}
	\label{fig7}
\end{figure*}

\begin{figure}[!t]
	\centering
	\includegraphics[scale=0.325]{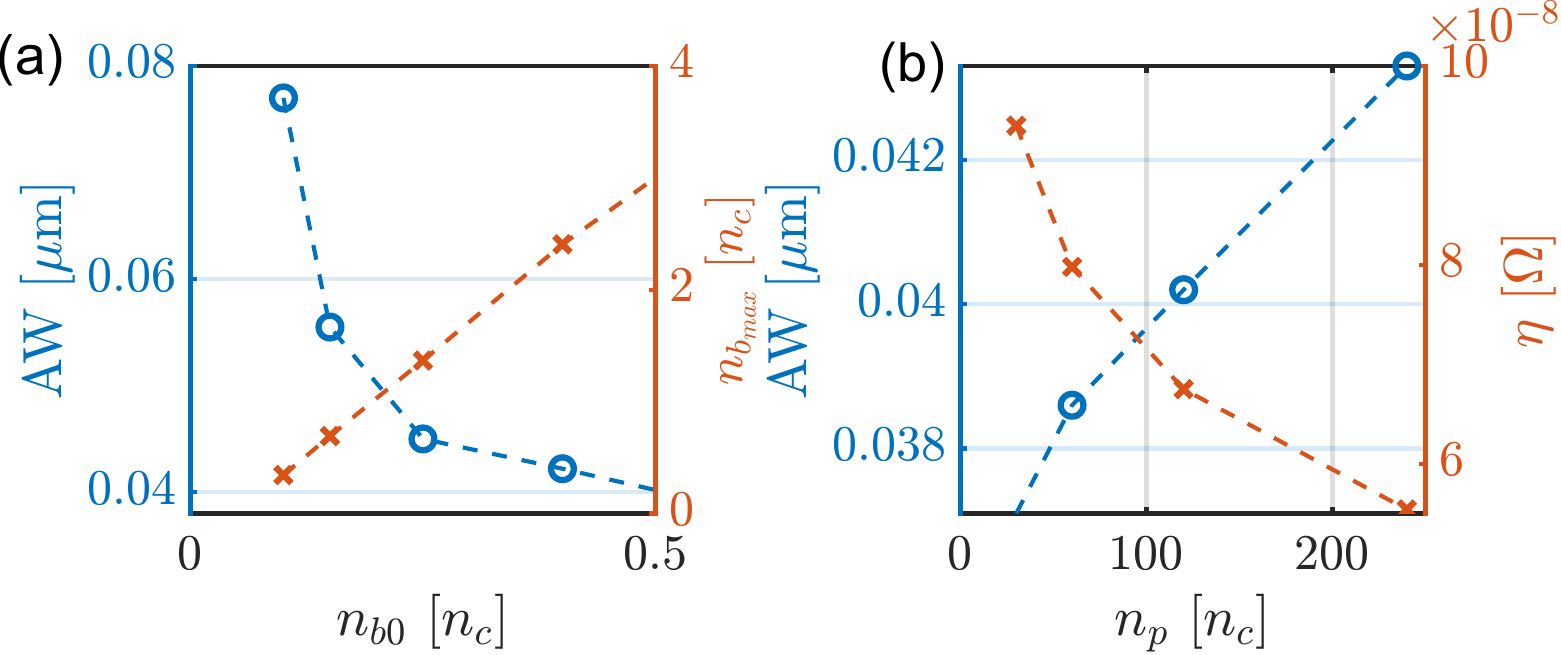} 	
	\caption{(a) Minimum AW and corresponding peak density under different beam densities. The blue dotted line and marker represent the minimum AW, the orange dotted line and marker indicate the corresponding peak density. (b) Minimum AW and corresponding peak density under different background plasma densities. The blue dotted line and marker represent the minimum AW, the orange dotted line and marker indicate the corresponding peak density. These results are obtained by fitting the transverse distribution of annular electron beams with Gaussian function.}
	\label{fig8}
\end{figure}

\section{Simulation Results}
   In order to confirm the scheme we proposed, we have performed 3D simulations by using LAPINS code \cite{Wu2023POP}. The parameters used in the simulation are the same as those in the theoretical section. The simulation box is $2\ \mathrm{\mu m}(x)\times 2\ \mathrm{\mu m}(y)\times 6$ or $10\ \mathrm{\mu m}(z)$, which is divided into $100 \times 100 \times 300$ or $500$ grids and is filled with homogeneous hydrogen plasma. The annular electron beam enter from the $z=0$ boundary along the $z$ direction, with twenty-seven microparticles of each species in each grid. Collision effects are considered in all 3D simulations.

   Figs.\ \ref{fig3}(a) and \ref{fig3}(b) show the theoretical and simulated transverse magnetic field for $\tau=15T_0$ in the case without OAM, respectively. The simulation results are consistent with the theoretical results. There are magnetic fields with opposite signs inside and outside the annular electron beam, and the Lorentz force of the magnetic field on the outside of the annular electron beam points to the inside of the ring, while the Lorentz force of the magnetic field on the inside of the annular electron beam points to the outside of the ring. It leads to the focusing of the annular electron beam. Fig.\ \ref{fig2} also shows that the amplitude and spatial distribution of azimuthal magnetic field obtained from the theory agree well with the simulation results. In Fig.\ \ref{fig4}, we draw distribution of electron beam density for $\tau=15T_0$. We can found that the annular electron beam is focused with a very thin AW, and the density of the electron beam is increased by six times. We also notice the electron beam is somewhat divergent because of over-focusing. At this time, by controlling the transport distance of electron beam in plasma, we can obtain an annular electron beam with appropriate AW.
   \par

   To see the transport characteristics of electron beams with OAM in plasma, we have carried out 3D simulations of annular electron beam with azimuthal velocity. Electrons carry counterclockwise azimuthal velocity and other parameters of the annular electron beam are in agreement with those presented in the previous theoretical section, and the background plasma properties remain consistent with the aforementioned simulations. In the plasma medium, it has been observed that the AW of the annular electron beam carrying OAM is gradually reduced because of the self-generated transverse magnetic field, but the annular electron beam diverges outward since there is no force acting to sustain its circular motion, as showed in Fig.\ \ref{fig5}(a). To control the transport of the electron beam, we implement a longitudinal magnetic field. Centripetal force provided by Lorentz force, we have
   	\begin{equation}
   		B_z q v_\theta=\gamma m_e\frac{v^2_\theta}{r_c},
   		\label{eq15}
   	\end{equation}
   	 where $\gamma$ represents the Lorentz factor, $v_\theta$ is the electron's lateral velocity, and $r_c$ denotes the electron's cyclotron radius. According to $p_\theta=2 r m_e c/\lambda$, we can deduce that the cyclotron radius of electrons is
   	 \begin{equation}   	 	
   	 	r_c =\frac{2m_ecr}{B_zq\lambda}
   	 	\label{eq16}
   	 \end{equation}
   	 In the simulations, we set the magnetic field to $1\times 10^4\ \mathrm{T}$, resulting in a cyclotron radius for the electron of approximately $0.34r$, which is smaller than the electron's transverse  position. Additionally, due to the uniformity of the magnetic field, the cyclotron periods of the electrons can be considered approximately equal, allowing the electrons on the same circumference to occupy the same cyclotron phase. The extended lines of the cyclotron radii for different electrons intersect at the center of the annular electron beam, ensuring that the topological structure of the electron beam remains unchanged, as shown in Fig.\ \ref{fig6}(a).  When the cyclotron radius $r_c>r$, the radius of the annular electron beam will increase and oscillate periodically within a certain range. In this range, the minimum radius corresponds to the initial radius of the annular electron beam, while the maximum radius is $2r_c-r$, as illustrated in Fig.\ \ref{fig6}(b).
   \par

   The azimuthal magnetic field generated when the annular electron beam carries OAM is shown in Fig.\ \ref{fig2} (red dotted line), it is only slightly different from that without OAM. It can be seen that the magnetic field of azimuthal with OAM is slightly reduced compared with that without OAM, which is consistent with the trend of theoretical results. Fig.\ \ref{fig5}(b) show the variation of the annular electron beam density for the case with OAM and the longitudinal magnetic field. The radius of the annular electron beam rapidly decreases under the influence of the longitudinal magnetic field. Meanwhile, the electron beam is focused due to the transverse focusing magnetic field and the longitudinal external magnetic field. However, when the annular electron beam is focused to about $0.3\ \mathrm{\mu m}$ in diameter, the electron beam begins to expand outward and remain a thin AW. Then the radius of annular electron beam varies periodically. This is due to the periodic nature of the electron beam's cyclotron motion. Fig.\ \ref{fig7} illustrates the motion state of electrons in the simulation, with the white arrow indicating the direction of electron motion. Additionally, the electron cyclotron phase and motion direction are represented by a red solid line and arrow for clarity. In this figure, segments (a) to (c) represent electrons in the $0$ to $\pi/2$ phase, (c) to (e) correspond to the $\pi/2$ to $\pi$ phase, (e) to (g) cover the $\pi$ to $3\pi/2$ phase, and (g) to (h) represent the $3\pi/2$ to $2\pi$ phase. The cycle then repeats; however, during this process, under the influence of the angular magnetic field, the AW of the annular electron beam remains relatively stable and no longer shows significant variation. According to Fig.\ \ref{fig7}, we can determine that the electron cyclotron motion period is approximately $19.67\ \mathrm{fs}$. Additionally, using the cyclotron formula under relativistic conditions, we calculate the electron cyclotron period to be approximately $22.58\ \mathrm{fs}$. This result further validates the robustness of our analysis. The cyclotron radius of the electrons is affected by the strength of the external magnetic field. This allows us to control the transverse radius of the annular electron beam accordingly.
   \par

\section{Discussion and Conclusion}
   Our theoretical analysis indicates that the factors affecting the AW of the annular electron beam include electron beam current intensity, temperature, and background resistivity. In Fig.\ \ref{fig8}(a), we show the variation of AW with different beam and background plasma densities. We observe that as the electron beam density increases, the AW of the annular electron beam decreases, aligning well with our theoretical predictions. Besides, an increase in background plasma density results in a slight increase in electron beam AW due to a decrease in background resistivity. However, because this change in resistivity is subtle, background density has a limited effect on the beam AW, as shown in Fig.\ \ref{fig8}(b). Therefore, the AW of the annular electron beam can be manipulated by varying the beam current intensity, temperature, and background resistivity.
   In the case of carrying OAM, the radius of the annular electron beam will be modified by the external longitudinal magnetic field. In our simulations, to expedite phenomenon observation and conserve computing resources, we utilize relatively limiting angular velocity parameters. In fact, the angular velocity and the required magnetic field are much smaller than those selected for our simulations. These parameters are also more feasible in the laboratory.
   \par
   In conclusion, the transport characteristics of annular electron beam in a plasma medium are studied theoretically and numerically. Through the rigid beam model, we deduce that there is an azimuthal magnetic field in the process of electron beam transport in the plasma medium. This magnetic field can alter the AW of the annular electron beam. In addition, it is considered that in some schemes for generating annular electron beams, the electron beams will carry OAM. Thus we have carried out numerical simulations of the transport of an annular electron beam with OAM in the plasma medium. Our theory and simulations indicate that the possession of OAM by the annular electron beam has minimal impact on the strength of the azimuthal magnetic field. However, we note distinct transport characteristics of the annular electron beam with OAM compared to those without OAM. Specifically, the former necessitates an external axial magnetic field to ensure its stable propagation in the plasma medium. Under the influence of this magnetic field, the radius of the annular electron beam can oscillate periodically, with the direction of change whether increasing or decreasing dependent on the field's strength. In this case, we can not only control the AW of the annular electron beam but also the radius of the annular electron beam. By this conclusion, we are expected to produce high-quality annular electron beams with controllable annular and radius. This can greatly broaden the application scene of annular electron beam.
\section{Acknowledgements}
   D.W. initiated this work. Y.C.L. performed the simulations and data analysis. Y.C.L. and D.W. drafted the manuscript. All the authors contributed to physical analysis. This work was supported by the National Natural Science Foundation of China (Grant No. 12075204, No. 11875235, and No. 61627901), the Strategic Priority Research Program of Chinese Academy of Sciences (Grant No. XDA250050500) and Shanghai Municipal Science and Technology Key Project (No. 22JC1401500). Dong Wu thanks the sponsorship from Yangyang Development Fund.
\bibliography{reference.bib}
\end{document}